\documentclass[aps,prb,twocolumn,floatfix,showkeys,showpacs,superscriptaddress]{revtex4}
\pdfoutput=1
\usepackage{latexsym} \usepackage{amsmath} \usepackage{dcolumn}
\usepackage{bm} \usepackage{graphicx} \usepackage{epsfig}
\graphicspath{{./Figures/}}

\usepackage{verbatim}
\usepackage{hhline}

\newcommand{\detailtexcount}[1]{%
  \immediate\write18{texcount -merge -sum -q #1.tex output.bbl > #1.wcdetail }%
  \verbatiminput{#1.wcdetail}%
}

\begin{document}


\title{Ab initio multiplet plus cumulant approach
for correlation effects in x-ray photoelectron spectroscopy}



\author{J. J. Kas} \affiliation{Dept.\ of Physics, Univ.\ of
Washington Seattle, WA 98195-1560}
\author{J. J. Rehr} \affiliation{Dept.\ of Physics, Univ.\ of
Washington Seattle, WA 98195-1560}
\affiliation{Department of Photon Science, SLAC National Accelerator Laboratory, Menlo Park, California 94025, USA}
\author{T. P. Devereaux} 
\affiliation{Department of Materials Science and Engineering, Stanford University, Stanford, California 94305, USA}
\affiliation{Stanford Institute for Materials and Energy Sciences, SLAC National Accelerator Laboratory, Menlo Park, California 94025, USA}


\date{\today}

\begin{abstract}
The treatment of electronic correlations in open-shell systems is among the most challenging problems of condensed matter theory.  Current approximations are only partly successful. Ligand 
field multiplet theory (LFMT) has been widely successful in describing intra-atomic correlation effects in x-ray spectra, but typically ignores itinerant states.
The cumulant expansion for the one electron Green’s function has been successful in
describing shake-up effects 
but ignores atomic multiplets. More complete methods, such as dynamic
mean-field theory can be computationally problematic.   Here we show that
separating the dynamic Coulomb interactions into  local and longer-range parts
with {\it ab initio} parameters yields a combined  multiplet plus cumulant approach that
accounts for both local atomic multiplets and satellite excitations. 
The approach is illustrated in transition metal oxides and explains the
 multiplet peaks, charge-transfer satellites and 
distributed background features observed in XPS experiment.
\end{abstract}

\pacs{71.15.-m, 31.10.+z,71.10.-w}
\keywords {Green's function, cumulant, GW, DFT, excited states}

\maketitle

The frontier of  an {\it in-silico} description of complex functional behavior of materials lies in a consistent description of their physics and chemistry over many length and time scales.
Achieving such a description has been a major challenge of computational materials science, whose success is ultimately assessed by the ability to describe 
both  ground and excited states, 
including non-equilibrium conditions \cite{martinreiningceperley}.  Computational methods   typically focus on 
delocalized band-like descriptions, as in Kohn-Sham density functional theory (DFT), or highly localized descriptions, as in ligand-field multiplet theory (LFMT) \cite{degrootbook,shirley2005} and quantum chemistry \cite{BAGUSReview,EOMCC-Bartlett}. However,  
more elaborate approaches such as dynamical mean field theory (DMFT) \cite{DMFT,biermann12} and embedding methods extended to clusters \cite{ghiasi_dmft,Hariki_2017} have been used to  
move away from  purely local or delocalized points of view.
Both aspects are important in x-ray spectroscopy,
which can provide atom-specific information about local coordination, valency, and excited states.
Quantitative approaches that treat both aspects  without adjustable parameters are  highly desirable.    This is the main goal of this work, where an {\it ab initio} approach combining a local LFMT model  with the non-linear cumulant Green's function  \cite{hedin99review,zhou2017,KasRC,Tsavala} is developed to treat both local- and longer-range correlations.

  Core-level x-ray photoemission spectroscopy (XPS) is a sensitive probe of correlation effects in excited state electronic structure. In particular, the XPS signal is directly related to the core-level spectral function $A_c(\omega)$, which  describes the distribution of excitations in a material. 
  The main peak in the XPS corresponds to the quasiparticle, while secondary features, i.e., satellites, correspond to many-body excitations. These satellite features are pure many-body correlation effects that  have proved difficult to calculate  from first principles in highly correlated materials.
  They can also have considerable spectral weight, comparable to that in the main peak and spread over a broad range of energies.     
  Many-body perturbation theory within the GW approximation is inadequate to treat these effects. While GW can give reasonably accurate core-level binding and quasi-particle energies  \cite{vanSetten,Golze}, the satellite positions and amplitudes are not well reproduced, even in relatively weakly correlated systems such as sodium and silicon. 
On the other hand, the cumulant expansion of the one-electron Green's function \cite{ND,langreth69,almbladh} has had notable success in predicting the quasi-bosonic satellite progressions  \cite{hedin99review,guzzo,guzzo14,lischner,ZhouAluminum,Caruso2020}, as well as the charge-transfer satellites observed in some correlated materials \cite{KRC,particlehole}. 
Nevertheless a quantitative treatment of the excitation spectrum of strongly correlated materials  demands more elaborate theories. Theories like  CI \cite{BAGUSReview}, coupled-cluster  \cite{eomcc1,eomcc2}, or model Hamiltonian methods fit to  DFT or GW calculations as in \textit{ab initio} LFMT \cite{haverkort2012,ikeno2011,singh_ailfmt,kruger_ailfmt},  can yield impressive results for the multiplet splittings   seen in XPS. However, they usually lack an adequate Hilbert space to account for  extended states and collective
excitations. Moreover, these theories typically include some adjustable parameters which may obscure the underlying physics.  Although dynamical processes such as charge-transfer excitations 
can be treated with  cluster LFMT or LDA+DMFT \cite{sipr_dmft,ghiasi_dmft}, the inclusion of higher energy excitations has remained numerically challenging \cite{biermann12}. 

In an effort to address these limitations, we introduce here an {\it ab initio} approach which combines a local multiplet model
that ignores charge transfer excitations, with a non-linear cumulant approximation for the core-Green's function. We dub the approach {\it Multiplet+C}, analogous to  other methods where the cumulant Green's function is added to treat satellite excitations \cite{hedin99review,guzzo,lischner,Aryasetiawan,KRC}. Our approach is advantageous computationally, both for its  simplicity in implementation  and its physical interpretation. 
For definiteess, we focus here on the $2p$ XPS of  transition metal oxides (TMOs). By separating the short- and long-ranged Coulomb interactions, the method yields an expression for the core spectral function $A_{2p}(\omega)=-(1/\pi){\rm Im}\, G_{2p}(\omega)$ given by
a convolution of the local model and cumulant spectral functions,
\begin{equation}
    A_{2p}(\omega) = A_{2p}^{\rm loc}(\omega) * A_{2p}^{C}(\omega).
\end{equation}
Thus each discrete local multiplet level is broadened by $A_{2p}^C(\omega)$, which accounts for shake satellites and an extended tail. As a consequence, our combined method treats both local  correlations and dynamical, more extended  excitations. Similar convolution methods have been used to add many-body excitations to correlated systems \cite{ajlee,calandra2012}. 
In the time-domain, the cumulant ansatz gives the Green's function as a product of the local Green's function and an exponential of the cumulant $C_{2p}(t)$, 
\begin{equation}
    G_{2p}(t) = G_{2p}^{\rm loc}(t)e^{C_{2p}(t)}.
\end{equation} 
Here $G_{2p}^{\rm loc}(t)$ is the trace over $2p$ single particle states of the atomic multiplet Green's function for our   local atomic model, and $C_{2p}(t)$ is the cumulant, wich is calculated in real-time \cite{KasRC}, and builds in  dynamic correlation effects. The above approximation was inspired by the work in Refs.\ \cite{biermann12,biermann16}, where a similar product of an atomic Green's function and cumulant spectral function was used to treat plasmon excitations in DMFT.   

To formalize this  approach, we define a separable model Hamiltonian $H=H^{loc} + H^{bos}$ in which the  localized 
system consists of a limited number of electrons (the 2$p$ and 3$d$ shells for example) interacting with the extended system via quasi-bosons that characterize the many-body excitations \cite{hedin99review}. Here local system is defined by a  many-body Hamiltonian,
\begin{equation}
    H^{\rm loc} = \sum_{i} \epsilon_{i} n_{i} + \sum_{i,j}[V^{\rm xf}_{i,j}c^{\dagger}_i c_j + c.c.] + \sum_{i,j,k,l} v_{ijkl}c^{\dagger}_ic^{\dagger}_j c_k c_l,
\end{equation}
where $V^{\rm xf}$ denotes the crystal field potential, $v$ the Coulomb interaction, and  the electron levels $\{i,j\}$ are limited to the 2$p$ and 3$d$ shells of a single atom. More generally this Hamiltonian could be extended to include ligands, as in cluster LFMT and CI calculations. 
Thus $H^{\rm loc}$ accounts for covalency effects on the multiplet levels, but ignores charge transfer satellites which are included via the cumulant. Additional details are given in the Supplementary Material.
Although this approximation yields  a simple solution of the full problem, it ignores possible
final-state effects of charge transfer on the local configuration, and instead keeps a fixed number of $d-$electrons in the local Hamiltonian. 

The quasi-boson Hamiltonian for the extended system including the coupling to the localized system is
\begin{equation}
    H^{\rm bos} = \sum_q \omega_q a^{\dagger}_q a_q + \sum_{qi} n_iV_i^{q}(a^{\dagger}_q + a_q),
\end{equation}
where $V_i^q$ are fluctuation potentials \cite{hedin99review}, and $n_i$ is the occupation of the hole-state $i\in 2p$.
If we now approximate the couplings $V^q_i$ to be independent of the multiplet-hole state $i$ of the localized system, the net coupling depends only on the total number of holes $N_h = \sum_i n_i$ in the $2p$ shell, which is equal to $1$ in the XPS final state.
Then the Hamiltonian $H^{bos}$ becomes  equivalent to that of Langreth \cite{langreth69}, which describes a system of bosons interacting with an isolated core-electron. Notably, this model can be solved using a cumulant Green's function, with a cumulant proportional to the density-density correlation function $\chi(q,q',\omega)$ and  
 yields a spectral function with a  series of satellites  corresponding to
bosonic excitations. The difference in our treatment is that the localized system has it's own set of eigenstates (the atomic-multiplet levels) once the 2$p$ hole appears, each with it's own bosonic satellites from the convolution with $A^C_{2p}(\omega)$

 Our {\it ab initio} calculations of the local LFMT model include extensions to account for strong correlation effects.
The local multiplet system defined by  $H^{loc}$ ignores charge-transfer coupling and depends on several parameters. The Slater-Condon parameters $F$ and $G$ are calculated using self-consistent radial wave functions which take covalency into account, averaged over the occupied $3-d$ states, and $2p$ core-level states from the modified Dirac-Fock atomic code of Desclaux \cite{ankudinov1996,zhou2017,desclaux} available within FEFF10 \cite{Feff10}.
We find that the calculated $F$ and $G$ values are typically reduced from free-atom values by a factor of 
about 0.7 to 0.8 due to covalency effects, consistent with other studies \cite{Chen,haverkort2012}.
The crystal field strengths $10Dq$ were estimated from the $T_{2g}-E_g$ splitting in the angular momentum projected densities of states from FEFF10, and aregiven by $0.8$ and $1.3$ eV  for Fe$_2$O$_3$ and MnO respectively, although the the spectra are not particularly sensitive to these values. Spin-orbit couplings were taken to be the atomic values \cite{haverkort_thesis}, which can also be obtained from the Dirac-Fock code in FEFF10.  We have verified that the resulting multiplet spectra  for  hematite obtained with this model agrees well  
 with the accurate cluster CI calculations of Bagus et al.\ \cite{bagus_hematite},
that also ignore charge-transfer coupling (shake satellites). Further details are reported in the Supplemental Material at [URL will be inserted by publisher].
 
The  cumulant Green's function is calculated with a cumulant $C_{2p}(t)$ analogous to that in the Langreth formulation, but obtained using a modified real-time TDDFT approach\cite{KRC}. Within the  Landau representation \cite{landau44}, 
\begin{align}
    C_{2p}(t) &= \int d\omega \frac{\beta(\omega)}{\omega^2}\left[e^{-i\omega t}+i\omega t - 1\right], \nonumber \\
    \beta(\omega) &= \omega \int d^3r {\rm Re}[V(r)\delta\rho({\bf r},\omega)].
    \label{eqn:cumulant}
\end{align}
Here $\delta\rho({\bf r},\omega)$ is the time-Fourier transform of the density fluctuations $\delta\rho(r,t)$ induced by the sudden appearance of the core-hole, and $V(r)$ is the $2p$ core-hole potential. In order to treat the strong core-hole effects in correlated systems, we include non-linear corrections to the cumulant, following Tzavala et al.\ \cite{Tsavala}. 
A  measure of correlation strength is given by the dimensionless satellite amplitude
$a=\int\,d\omega \beta(\omega)/\omega^2 = -\ln Z$
where $Z$ is the renormalization constant. Note that $a$ is sensitive to the behavior of $\beta(\omega)$ near $\omega=0$.
To account for the energy gap in TMOs, which is not well treated in TDDFT and  affects the asymmetry of the quasi-particle peak, we set the linear part of the cumulant kernel $\beta(\omega)$ at low
$\omega$ to zero. This is in contrast with the use of the scissor operator, which shifts unoccupied states  uniformly to higher energies by the gap correction; however, recent calculations show that the shake up satellite energies are not affected by the gap correction \cite{ghiasi_dmft}.
\begin{figure}[tp]
    \centering
    \includegraphics[width=0.9\columnwidth]{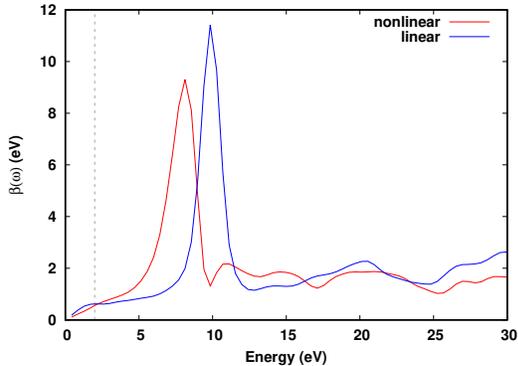}
    \caption{Quasi-boson excitation spectrum  $\beta(\omega)$ of Eq.\ (\ref{eqn:cumulant}) for hematite calculated within linear- (blue) and  non-linear (red) response using the RT-TDDFT approach \cite{KasRC}. The main peak corresponds to the charge-transfer excitation energy, which is red-shifted by non-linear response,  and the tail corresponds to a broad background. The grey dashed line shows the energy below which the spectrum was set to zero to correct the gap. 
    \label{fig:nlspecfn}}
\end{figure}

As illustrative examples, we apply   the {\it Multiplet+C} approach to the 2$p$ XPS of $\alpha$-Fe$_2$O$_3$ (hematite) and
MnO. Both the Fe and Mn sites in these systems are octahedrally coordinated by O, although there is distortion from octahedral symmetry in Fe$_2$O$_3$. In addition, the nominal oxidation state is different in the two materials, i.e., Fe$^{3+}$ and Mn$^{2+}$ in the ground state. Nevertheles, their multiplet structure is similar, since both metal atoms are nominally $d^5$. At room temperature, MnO is paramagnetic, while Fe$_2$O$_3$ is antiferromagnetic, but the magnetic structure seems to have little affect on the shake satellites in Fe$_2$O$_3$ \cite{Hariki_2017}.

Our results for the cumulant kernel $\beta(\omega)$ for $\alpha-$Fe$_2$O$_3$ (hematite) are shown in 
Fig.\ \ref{fig:nlspecfn}.   Note that the
 non-linear corrections further broaden and red-shift the main satellite in the direction of the -9 eV peak in the experimental XPS. Thus our calculation of $\beta(\omega)$  differs  from the conventional treatment  based on linear response and the GW approximation,
$\beta^{GW}(\omega) = (1/\pi) |{\rm Im}\, \Sigma^{GW}(\omega+\epsilon_c)|$ \cite{hedin99review}. Within our simplified model Hilbert space, there are no excitations due to charge transfer from the localized system to the surroundings or vice versa. Consequently the spherical contributions to the direct interactions $F^{0}_{pd}$ and  $F^{0}_{dd}$, only contribute  to overall static shifts in the spectrum but not satellite structure. Thus in order to avoid double counting in the calculation of shake-up or charge-transfer satellites, we only use the spherical part of the Coulomb interaction when calculating the density response $\delta\rho(t)$ to the suddenly created core-hole at $t=0$.
\begin{figure}[t]
    \centering
    \includegraphics[width=0.9\columnwidth]{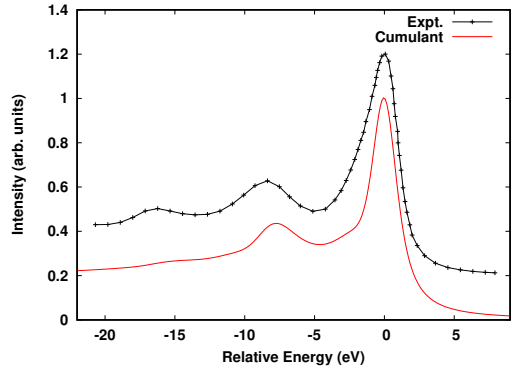}
    \caption{Spectral function $A_{1s}(\omega)$
    for Fe$_2$O$_3$ compared with the 1s XPS \cite{miedema2015}, both normalized by the main peak height. Although the position and strength of the main satellite calculated with the non-linear cumulant spectral function are in reasonable agreement with experiment, the calculated satellite at -16 eV is too weak. 
     \label{fig:qp_sat_spfcn}}
\end{figure}
The spectral function associated with the cumulant
$A^C_{1s}(\omega)\approx A^C_{2p}(\omega)=(-1/\pi) {\rm Im}\, \mathcal{F}\left\{\exp[C(t)]\right\}$,
where $\mathcal {F}$ denotes a Fourier transform, is directly related to the 1s XPS which does not have atomic multiplet splitting, as shown in Fig.\ \ref{fig:qp_sat_spfcn}.
\begin{figure}[b]
    \centering
    \includegraphics[width=0.9\columnwidth]{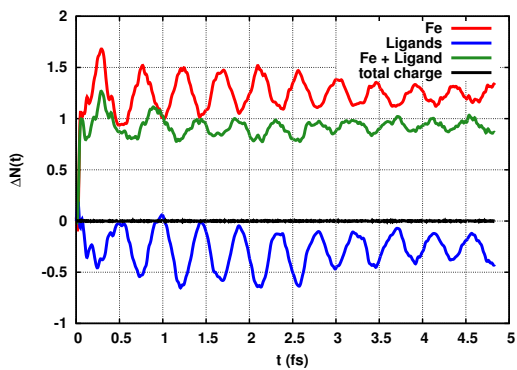}
    \caption{Fluctuation in Mulliken counts $\Delta N(t)$ on the central Fe atom (red), the $6$ O-ligands (blue), the sum of Fe and O-ligands (green), and the total (black) (which is zero to high accuracy) in hematite. Note that these counts oscillate with period $\sim$ 0.46 fs, corresponding to the charge-transfer frequency $\omega_{CT} \sim 9$ eV.   \label{fig:mulliken}}
\end{figure}
To better  characterize  the nature of the satellite excitations, it is useful to analyze the induced charge associated with the main shake satellite at -9 eV,  following creation of the $2p$ core hole. Our calculations of the fluctuations in the Mulliken charges (Fig.\ \ref{fig:mulliken}) vs time for hematite on the central Fe atom (red), on the $6$ O-ligand atoms (blue), as well as the sum of Fe and O-ligands (green). Within a fraction of a femtosecond,  the electron count on the Fe increases by ~1., then oscillates between 1.\ and 1.5 at a  frequency $\omega_{CT} \sim 9 $ eV, corresponding to charge-transfer fluctuations.
In contrast, the oscillations in the 
  O-ligand count are $180^o$ out of phase, indicating  substantial charge transfer between metal and ligand. Note, however, that the sum of ligand and metal counts (green) contains sizable residual oscillations, 
  indicating some  charge transfer from outer shells. This sum retains most of the initial increase seen in the Fe atom, suggesting that the transient screening in the first fraction of a femtosecond is collective in nature. Finally, although not shown, the oscillations are dominated by the minority spin channel on the Fe atom. This is not surprising, as the majority spin channel has only a small number of unoccupied $d$-states.
\begin{figure}[ht]
    \centering
    \includegraphics[width=0.9\columnwidth]{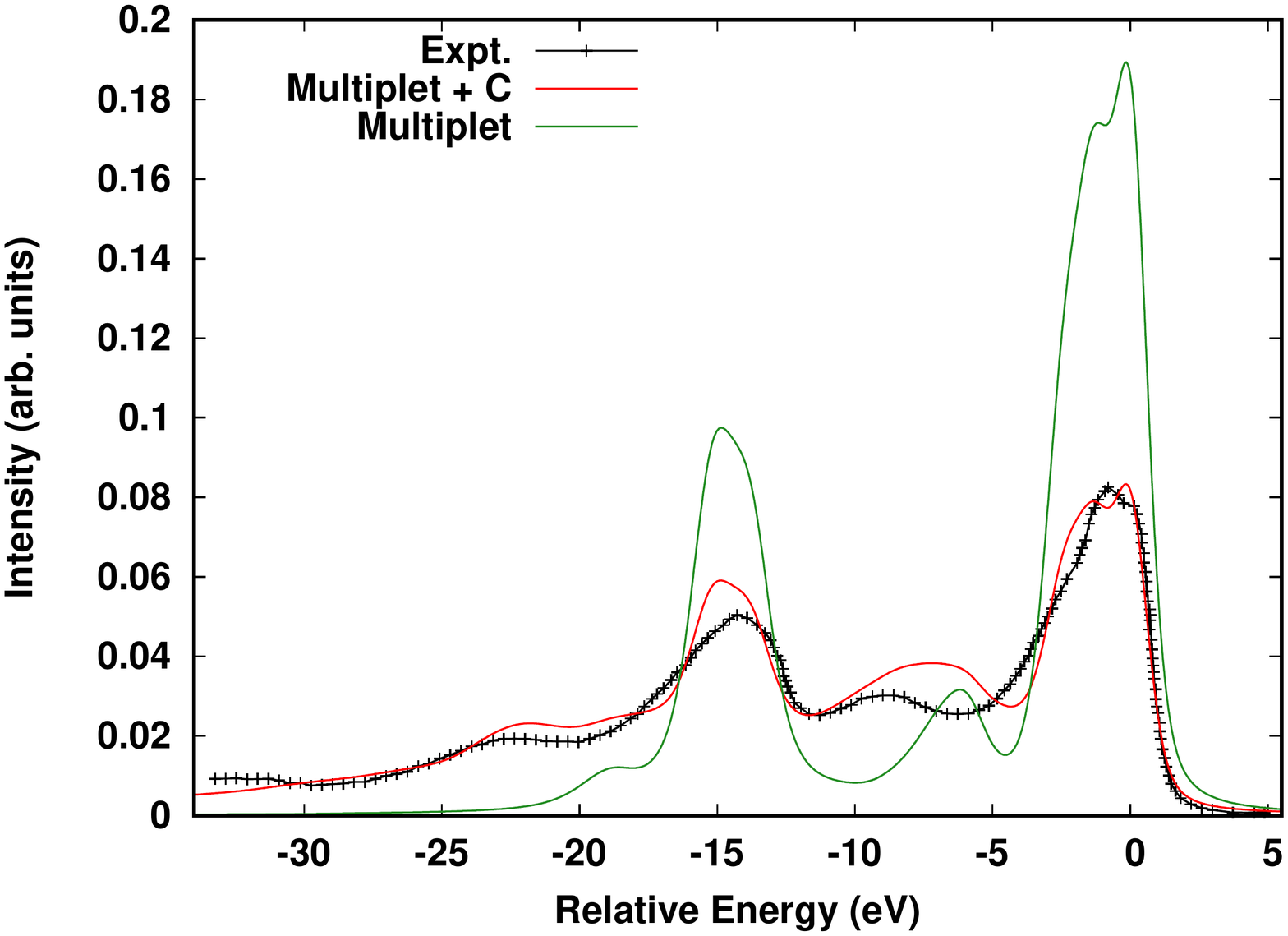}
    \includegraphics[width=0.9\columnwidth]{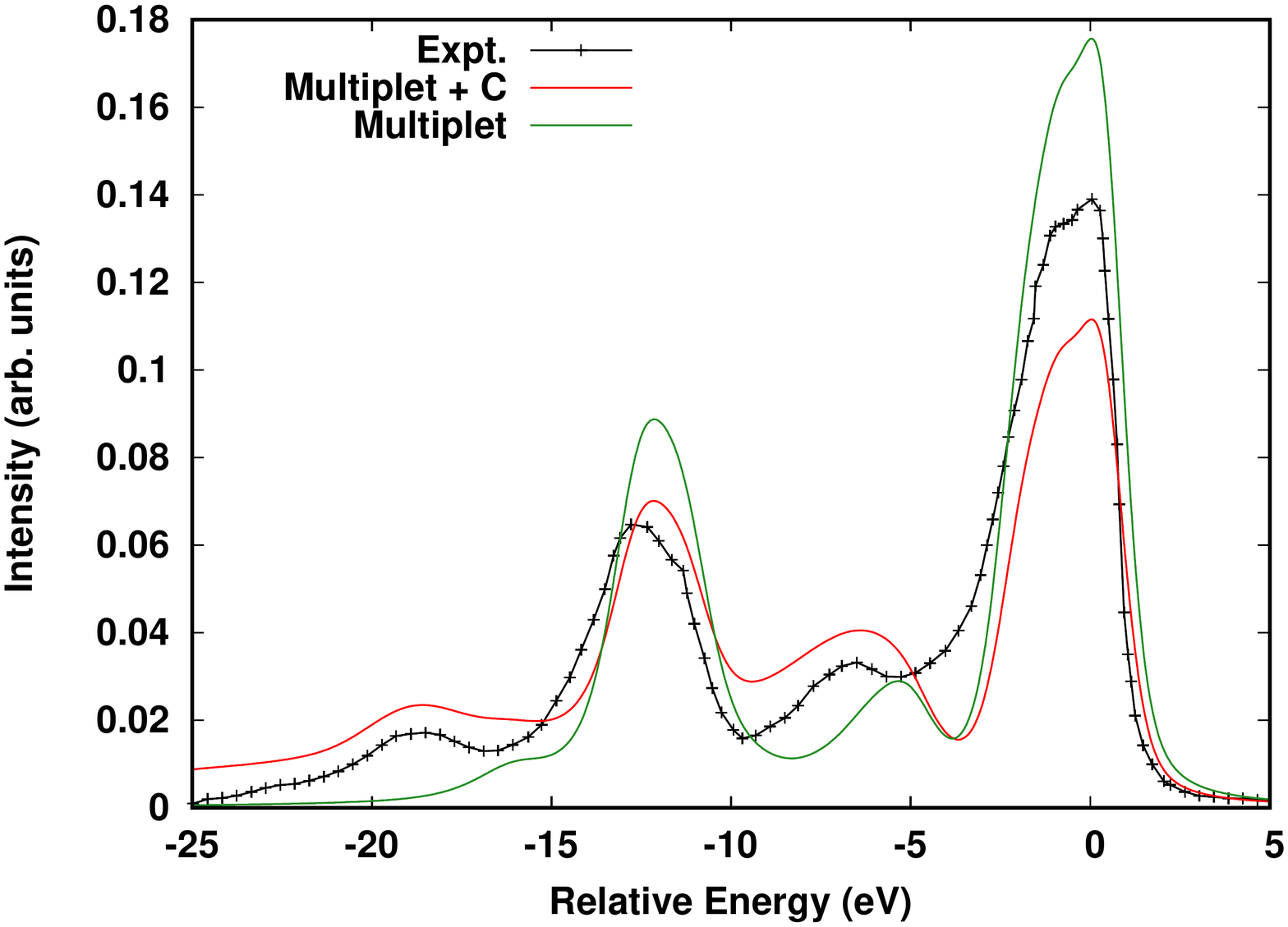}
    \caption{Comparison of the area normalized $2p$ XPS of: (Top) $\alpha-$Fe$_2$O$_3$ from experiment digitally reproduced from Ref.\ \cite{bagus_hematite} (black crosses)  with the {\it Multiplet+C} approach (red), and atomic multiplet only (green), and 
    (Bottom) Similar results for the XPS of MnO compared to  experiment (black crosses) \cite{Hariki_2017}. 
    Note that the multiplet-only calculations have relatively weak multiplet features $\sim 6$ eV below the 2p$_{3/2}$ main line which are not at the correct energy to explain the shake peaks. 
    \label{fig:femnxps}
    }
\end{figure}




Finally, Fig.~\ref{fig:femnxps} shows our results for the $2p$ XPS spectra from {\it ab initio} {\it Multiplet+C} approach,  our {\it ab initio} local atomic multiplet-only  model, and experiment for both hematite and MnO.  
 The most noticeable differences between the spectra with and without the cumulant spectral function
 are the broad satellites  roughly 7 eV and 9 eV below each of the main (2$p_{1/2}$ and 2$p_{3/2}$) peaks in MnO and Fe$_2$O$_3$ respectively. Upon convolution with the multiplet spectral function, they yield replicas of the local spectra at lower energies, with an energy splitting corresponding to the peak in the bosonic excitation spectrum $\beta(\omega)$.  In contrast the multiplet-only spectra have weak multiplet features at about -6 eV and no satellite below the 2p$_{1/2}$ peak or any substantial background intensity. While LDA+DMFT  calculations \cite{Hariki_2017} also yields comparable agreement for the main satellite, those results used some adjustable parameters. Moreover,
  there is a long-tail extending well beyond the satellites that contributes substantial intensity underneath the $2p_{1/2}$ main peaks, consistent with the background structure in $\beta(\omega)$ in 
  Fig.~\ref{fig:nlspecfn} and in experiment.  These properties reflect the different time-scales and dynamic correlation effects involved in the local and shake-up processes that are missing in conventional atomic multiplet models. 

 In conclusion, we have developed an {\it ab initio}
 {\it Multiplet+C} approach that treats both short- and longer-ranged excited state correlation effects in open-shell systems. 
The approach yields XPS spectra  for the full system as a convolution of the local multiplet spectrum
 and the non-linear cumulant spectral function for the extended system.
In this sense, the {\it Multiplet+C} approach is {\it doubly dynamic}, as in the DMFT+cumulant approach \cite{biermann12} and provides an attractive alternative to such methods. 
 Applications to Fe$_2$O$_3$ and MnO yield XPS spectra that agree reasonably well with experiment. 
 In contrast to  CI-based LFMT \cite{bagus_hematite}, Cluster-LFMT, or LDA+DMFT \cite{ghiasi_dmft}, our approach simplifies both the calculations and physical interpretation of the large-shake excitations and double-excitations in terms of density fluctuations induced by the suddenly turned on core-hole. Moreover, the approach  yields an {\it ab initio} treatment of the atomic muliplet spectra 
  and the broad background observed in experiment.
  Given the simplicity of the non-linear TDDFT approach and the approximate bosonic coupling, various improvements are desirable, especially for more strongly correlated systems like NiO. For example, better treatments of the density response and non-local screening corrections, as well as the effect of charge-transfer on the local model will likely be necessary for these systems \cite{Hariki_2017,ghiasi_dmft}.


Acknowledgments:  We thank P.\ Bagus, F.M.F.\ de Groot,  M.\ Haverkort, A.\ Hariki, and J.\ Lischner, L.\ Reining,   E.\ Shirley, and T.\ Fujikawa for helpful comments.  
This work was developed with support from the Theory Institute
for Materials and Energy Spectroscopies (TIMES) at SLAC, funded by the U.S. DOE, Offce of Basic Energy Sciences,
Division of Materials Sciences and Engineering under
contract EAC02-76SF0051.

\bibliography{MultiCumulant-resub}

\appendix

\section{ Multiplet+C Green's function}
A summary of the derivation of our combined {\it Multiplet + C} Green's function is as follows: Assuming smooth transition matrix elements, the $2p$ XPS is given by the partial trace over the states of the $2p$ shell of the Green's function, i.e., $I_{2p}^{\rm XPS}(\omega) \propto -({1}/{\pi}){\rm Im}\,G_{2p}(\omega) = -({1}/{\pi})\sum_{i\in 2p}{\rm Im}\, G_{ii}(\omega)$, where $G_{ii}(\omega)$ is the Fourier transform of the Green's function in the time-domai,  
\begin{equation}
    G_{ii}(t) = -ie^{-i E_0 t}\langle 0|c_{i}^\dagger e^{i H t}c_i|0\rangle. 
\end{equation}
Here $|0\rangle$ is the ground state of the system with energy $E_0$, and the Hamiltonian $H$ is separated into local and extended parts, i.e.,  
\begin{equation}
    H = H^{\rm loc} + H^{\rm bos}, 
\end{equation}
where $H^{\rm loc}$ describes a limited set of localized electronic states, and $H_{\rm bos}$ describes bosonic or quasi-bosonic excitations, i.e., plasmons, charge-transfer excitations, phonons, of the system etc.,
\begin{align}
    H^{\rm loc} &= \sum_i \epsilon_i n_i + \sum_{i,j}[V_{i,j}^{\rm xf}c_{i}^\dagger c_j + c.c.] + \sum_{i,j,k,l}v_{ijkl}c_{i}^\dagger c_{j}^\dagger c_k c_l, \nonumber \\
    H^{\rm bos} &= \sum_q \omega_q a_{q}^\dagger a_q + \sum_{q,i\in 2p}n_i V_i^q(a_{q}^\dagger + a_q).
\end{align}
Here $V_i^q$ are the fluctuation potentials, which couple the $2p$ hole state $i$ to the bosons labeled by an index $q$, $c_{i}^{\dagger}/c_{i}$ are electron creation/annihilation operators, and $a_{q}^\dagger/a_q$ are boson creation/annihilation operators. If we assume that $V_i^q = V^q$ is independent of the hole state, the boson Hamiltonian becomes,
\begin{equation}
    H^{\rm bos} = \sum_q \omega_q a_q^\dagger a_q + \sum_q N_h V^q (a_q^\dagger + a_q),
\end{equation}
where $N_h$ is the total number of $2p$ holes, equal to $0$ in the ground state, and $1$ in the XPS final state. Now the entire problem simplifies since $H^{\rm loc}$ and $H^{\rm bos}$ commute in the XPS final state, and the Green's function becomes,
\begin{align}
    G_{2p}(t) = -ie^{-i E_0 t}\sum_{i\in 2p}\langle 0|c_{i}^\dagger e^{i H^{\rm loc} t}e^{i H^{\rm bos} t}c_i|0\rangle.
\end{align}
In addition, the ground state is the zero hole, zero boson state, and can be written $|\Psi_0\rangle|\nu_0\rangle$, so the Green's function becomes separable,
\begin{align}
    G_{2p}(t) &= iG_{2p}^{\rm loc}(t)G^{\rm bos}(t), \nonumber \\
    G_{2p}^{\rm loc}(t) &= -ie^{-i E_0 t}\sum_{i\in 2p}\langle \Psi_0|c_i^\dagger e^{i H^{\rm loc} t}c_i|\Psi_0\rangle, \nonumber \\
    G^{\rm bos}(t) &= -i\langle \nu_0|e^{i H^{\rm bos} t}|\nu_0\rangle = -ie^{C(t)}.
\end{align}
The local Green's function can be calculated via exact diagonalization or iterative inversion schemes such as Lanczos. The Green's function of the bosons is identical to that of Langreth for an isolated core-electron interacting with bosons, and can be found analytically \cite{langreth69}. Finally, taking the Fourier transform, the XPS intensity is given by a convolution,
\begin{equation}
    I_{2p}^{\rm XPS}(\omega) \propto A_{2p}(\omega) = A^{\rm loc}(\omega)*A^{\rm bos}(\omega).
\end{equation}

\section{{\it Ab initio} Multiplet Model }
\begin{figure}[ht]
    \centering
    \includegraphics[width=1.0\columnwidth]{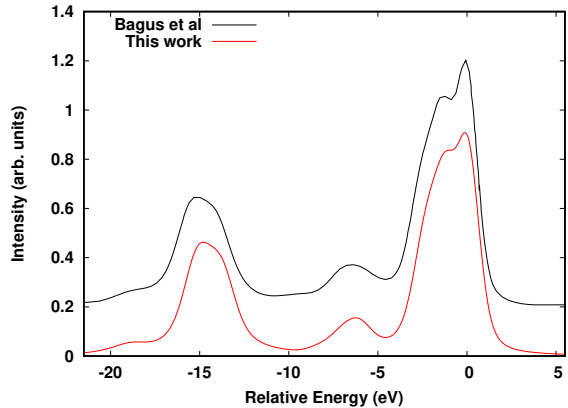}
    \caption{Comparison of the LFMT spectral funcions calculated using the CI-LFMT calculations of Bagus et al.\ (red)\cite{bagus_hematite} and  the atomic LFMT calculations using the calculated Slater-Condon parameters of this work (green).} \label{fig:multcomp}
\end{figure}

Ligand-field multiplet theory (LFMT) calculations were carried out with the WebXRS code \cite{wang2010,Devereaux2021}.    {\it Ab initio} values of the parameters were obtained using the updated FEFF10 code.
The Slater-Condon parameters $F$ and $G$ were obtaned using the radial wavefunctions available within the FEFF10 package \cite{feff10}, which build in covalency effects. For the core-levels, the Dirac-Fock atomic wavefunctions were used, while the valence wavefunctions were found using the self-consistent local density approximation, and were averaged over the occupied $d$-levels. These wavefunctions were normalized to the Norman sphere, and the radial integrals defining the Slater-Condon parameters were calculated with the Norman radius as an upper bound. The calculated Slater-Condon parameters are observed to be reduced by a factor of about 0.7 to 0.8 from their free-atom values (see table \ref{tbl:mult_par}.  The spin-orbit splitting was taken to be the atomic value as calculated in Ref.\ \cite{haverkort_thesis}, which agrees well with that that obtained from the Dirac-Fock code in FEFF10. These values also agree very well with those obtained with Wannier orbitals \cite{haverkort2012}.
  This approach 
  yields a multiplet spectrum that closely matches the first principles CI approach for an FeO$_6$ cluster calculated by Bagus et al.\cite{bagus_hematite}, as shown in Fig.~\ref{fig:multcomp}. The values of all parameters necessary to define the local atomic multiplet Hamiltonian are shown in Table~\ref{tbl:mult_par}.
  \begin{table}[ht]
\caption{ Parameters used in our {\it ab initio} LFMT calculations for Fe$^{3+}$ of Fe$_2$O$_3$ and Mn$^+$ in MnO. The Slater-Condon and crystal field parameters were calculated using the FEFF10 code (see text). For comparison free atomic values are also shown, which are typically reduced by a  factor of about 0.7 to 0.8.} 
\begin{center}
\begin{tabular}{ || c || c | c | c | c | }
\hline
& Fe$_2$O$_3$ & \% At. & MnO & \% At. \\
\hhline{#=#=|=|=|=|}
\hline
$F^2_{dd}$ & $10.1$  & $73$ & $9.4$ & $77$ \\ 
\hline
$F^4_{dd}$ & $5.8$ & $67$ & $5.5$ & $72$ \\
\hline
$F^2_{pd}$ & $5.7$ & $70$ & $5.2$ & $74$ \\
\hline
$G^1_{pd}$ & $4.1$ & $67$ & $3.7$ & $71$ \\
\hline
$G^3_{pd}$ & $2.3$ & $65$ & $2.0$ & $69$ \\
\hline
$10Dq$ & $0.8$ & & 1.3 & \\ 
\hline
$\zeta_{2p}$ & $8.2$ & & $6.8$ &\\ 
\hline
$\zeta_{3d}$ & $0.1$ & & $0.06$ &\\
\hline
\end{tabular}
\end{center}
\label{tbl:mult_par}
\end{table}

\section{ Real-time Cumulant}
{\it Ab initio} real-time TDDFT calculations of the cumulant $C(t)$ were carried out using the procedure described in Ref.\ \cite{KRC} using the real-time extension of the SIESTA code \cite{takimoto2007} with the PBE generalized gradient functional \cite{PBE}. The default double-zeta plus polarization (DZP) basis set was used for both Fe and O atoms in Fe$_2$O$_3$ and for both Mn and O atoms in MnO. In order to limit the interaction of the core-holes, a 3x3x3 supercell consisting of 270 atoms was used, which was found to produce converged spectra. The pseudopotentials were calculated using the ATOM pseudopotential code \cite{ATOM} with parameters taken from Rivero et al.\ \cite{rivero}.\\  

\section{Broadening of the spectra}

In order to account for experimental broadening and coupling to phonons, 
all spectra were broadened with Voigt functions using a Gaussian full-width 
at half-max (FWHM) of $1.2$ eV . To account for finite lifetime effects, the Lorentzian broadening was set to $0.41$ eV FWHM for the $2p_{3/2}$ 
portion of the spectrum, above $\approx -10$ eV, and increased to $1.14$ eV below that to account for the larger lifetime width of the $2p_{1/2}$ states. For 
the $1s$ spectrum, the Lorentzian broadening was set to $1.8$ eV FWHM.

\ifnum 1>1 {
%
  \immediate\write18{texcount -1 -sum -merge -q MultiCumulant.tex output.bbl > MultiCumulant-words.sum }%
  \input{MultiCumulant-words.sum} words%

  \immediate\write18{texcount -1 -sum -merge -char -q MultiCumulant.tex output.bbl > MultiCumulant-chars.sum }%
  \input{MultiCumulant-chars.sum} characters (not including spaces)%

\detailtexcount{MultiCumulant}
} \fi

\end{document}